\documentclass[aps,prd,twocolumn,groupedaddress,showpacs,letterpaper]{revtex4}
\usepackage{graphicx}

\newcommand{\delete}[1]{}

\newcommand{\be}{\begin{equation}}
\newcommand{\ee}{\end{equation}}

\def\beq{\begin{equation}}
\def\eeq{\end{equation}}
\def\bea{\begin{eqnarray}}
\def\eea{\end{eqnarray}}
\def\ba{\begin{array}}
\def\ea{\end{array}}

\begin{document}


\title{Improved constraints on non-Newtonian forces at 10 microns}


\author{Andrew A. Geraci$^1$}
\email[]{aageraci@nist.gov}
\thanks{Present address: National Institute of Standards and Technology,
Boulder, CO 80305}
\author{Sylvia J. Smullin$^1$}
\thanks{Present address: Makani Power, Inc. Alameda, CA 94501}
\author{David M. Weld$^1$}
\thanks{Present address: Department of Physics, Massachusetts
Institute of Technology , Cambridge, MA 02139}
\author{John Chiaverini$^1$}
\thanks{Present address: Los Alamos National Laboratory, Los Alamos, NM 87545}
\author{Aharon Kapitulnik$^{1,2}$}
\affiliation{$^1$Department of Physics, Stanford University, Stanford, CA
94305}
\affiliation{$^{2}$Department of Applied Physics, Stanford
University, Stanford, CA 94305}


\date{\today}

\begin{abstract}
Several recent theories suggest that light moduli or particles in ``large''
extra dimensions could mediate macroscopic forces exceeding gravitational
strength at length scales below a millimeter. Such new forces can be
parameterized as a Yukawa-type correction to the Newtonian potential of
strength $\alpha$ relative to gravity and range $\lambda$.  To extend the
search for such new physics we have improved our apparatus utilizing
cryogenic micro-cantilevers capable of measuring attonewton forces, which
now includes a switchable magnetic force for calibration.  Our most recent
experimental constraints on Yukawa-type deviations from Newtonian gravity
are more than three times as stringent as our previously published results,
and represent the best bound in the range of $5-15$ $\mu$m, with a $95\%$
confidence exclusion of forces with $|\alpha|$ $>$ 14,000 at $\lambda=10$ $
\mu$m.
\end{abstract}

\pacs{04.80.Cc}

\maketitle


\section{Introduction} A number of theories of physics beyond the Standard
Model suggest that new physics related to
 gravity may appear at sub-millimeter length scales.  For example, light moduli from string theory \cite{sg,add97} or
 exotic particles in ``large'' extra dimensions \cite{add1,iadd,add3} can mediate forces exceeding gravitational
 strength that can be observed in a tabletop experiment.  Such phenomena can be parameterized as a Yukawa-type
 correction to the Newtonian potential of strength relative to
 gravity $\alpha$ and range $\lambda$.  For two masses $m_1$ and $m_2$
 separated by distance $r$, the gravitational potential is modified to \be
 V(r) = -G_N \frac{m_1 m_2}{r}(1+\alpha e^{-r/\lambda}). \ee
 To search for such new forces, we have improved our cryogenic apparatus
utilizing silicon micro-cantilevers with attonewton force sensitivity
\cite{stanford1,stanford2}, which now includes a
 magnetic method for force calibration.   The cantilever is loaded with a rectangular gold prism fabricated by
focused-ion-beam milling that serves as a test mass for the experiment. The
driving (source) mass is moved horizontally beneath the cantilever at a
nominal vertical face-to-face separation of 25 $\mu$m. The force between
the masses is deduced from the displacement of the cantilever as measured
by a fiber-coupled laser interferometer.  We perform the measurement at the
cantilever resonance frequency, typically of order 300 Hz, while the
mechanical driving motion occurs at a sub-harmonic, typically one-third.
This is achieved by implementing a density modulation in the drive mass,
consisting of alternating gold and silicon sections.

For the magnetic calibration, a Co/Pt multi-layer film is deposited on the
test mass. The permanent magnetic moment couples to a magnetic field
gradient produced by current flowing across the meandering gold sections in
the drive mass device. The current is turned off for the Yukawa-force
search. A shield composed of high-magnetic-permeability material encloses
the cryostat to prevent the Earth's field from magnetizing the drive mass.
The amplitude and phase of a magnetic or Yukawa signal will change in a
predictable way as we vary the equilibrium position of the drive mass
oscillation. We utilize this scanning technique as an additional handle to
distinguish a signal from background forces. In this paper, the newest data
and error analysis are described, and the latest experimental constraints
on Yukawa-type deviations from Newtonian gravity are presented. Finally,
future directions are discussed.

\section{Experimental Setup}\label{expsetup}

\begin{figure}\begin{center}
\includegraphics[width=1.0 \columnwidth]{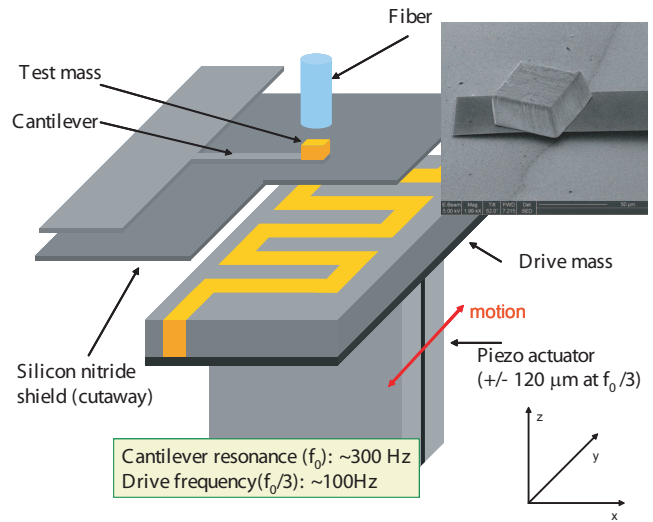}
\caption{(color online). A schematic (not to scale) showing the cantilever,
test and drive masses, shield, and piezoelectric bimorph actuator.  Figure
is adapted from Ref. \cite{jacthesis}. Not shown are the metallization of
the shield and drive mass ground plane.  A Fabry-Perot cavity is formed
between the bottom of the optical fiber and top of the test mass and is
used to interferometrically measure the cantilever displacement. The
coordinate axes shown in the lower inset are used throughout the paper to
describe the orientation of the drive mass and test mass. Upper inset: SEM
micrograph of FIB-fabricated test mass cut from 27 $\mu$m thick gold foil
attached to cantilever. \label{schematic}}
\end{center}
\end{figure}

Many of the details of the experimental probe and vacuum cryostat are
described in previous work \cite{sjsthesis,stanford1,stanford2}.  The
single-crystal silicon cantilevers are $250$ $\mu$m long, $50$ $\mu$m wide,
$0.3$ $\mu$m thick, and have a spring constant $k$ of approximately 0.0062
N/m. The thermal-noise-limited force sensitivity at 10 K is approximately
$200$ aN/Hz$^{1/2}$ for a typical low-temperature quality factor $Q$ of
80,000. The displacement of the cantilever beam is measured by using a
fiber-coupled laser interferometer where a Fabry-Perot cavity is created
between the end of the fiber optic and the cantilever loaded with a test
mass \cite{rugarintf}. The force $F$ on the cantilever is deduced from the
measured displacement $z$, which is enhanced on-resonance by the large
quality factor: $z = FQ/k$. The 1.5 $\mu$g test masses measure $54 \times
54 \times 27$ $\mu$m$^3$ and are cut from the edge of a $27$ $\mu$m thick
gold foil using a $20$ nA gallium focused-ion-beam (FIB). This technique
produced masses with more regular shapes than those used in previous
versions of the experiment \cite{stanford2} . A SEM image of a FIB-cut mass
attached to a cantilever is shown in the inset of Fig. \ref{schematic}. The
improvement allows a flatter surface to be presented to the drive mass,
allowing more force sensitivity, and results in an improved interferometer
signal. The driving (source) mass is mounted on a piezoelectric bimorph
which actuates in the $y$-direction (as shown in Fig. \ref{schematic})
beneath the cantilever with an amplitude of $\sim 120$ $\mu$m about its
equilibrium position. The drive mass consists of alternating 100
$\mu$m-wide bars of gold and silicon that are approximately 100 $\mu$m deep
and 1 mm long.  As the bimorph oscillates with a sinusoidal motion, a
time-varying gravitational force is exerted on the cantilever which can
occur at the driving frequency of the bimorph as well as at higher
harmonics, depending on the drive amplitude and spatial orientation of the
test and drive masses. Simulation indicates a maximal gravitational
coupling at the third harmonic of the drive frequency for a bimorph
amplitude of $133$ $\mu$m, with only a slight reduction ($\sim 10 \%$) at
120 $\mu$m \cite{andythesis}. A gold-coated silicon nitride shield membrane
separates the cantilever from the drive-mass and provides attenuation of
electrostatic and Casimir background forces from the oscillating drive
mass. The drive mass is covered with a smooth 1 $\mu$m-thick plane of
silicon, followed by aluminum oxide and gold to further suppress
modulations in electrostatic or Casimir forces associated with the sections
of alternating density. A schematic (not to scale) showing the cantilever,
test and drive masses, shield, and piezoelectric bimorph actuator is shown
in Fig. \ref{schematic}.  A piezoelectric stack actuator is also included
near the base of the cantilever to allow tests with deliberate excitation
and to facilitate interferometer offset control and characterization. The
voltage from the signal photodiode in the interferometer and the voltage
applied to the bimorph actuator are recorded at 10 kHz on a data
acquisition (DAQ) device connected to a PC. The cantilever signal (at the
third harmonic of the drive frequency) is averaged with respect to the
phase of the drive signal. A series of time records is collected for each
data point. An FFT of the interferometer data is performed to determine the
amplitude and phase at the third harmonic of the drive signal.

\begin{figure}\begin{center}
\includegraphics[width=1.0 \columnwidth]{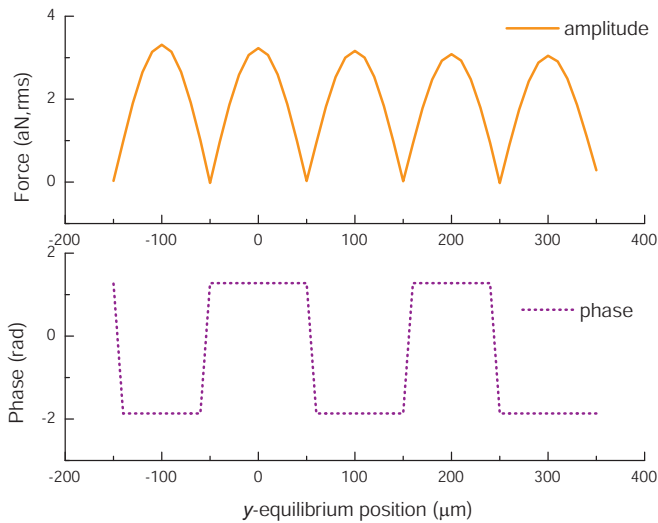}
\caption{(color online). Calculated vertical component of the Yukawa
($\alpha=1000,\lambda=18 \mu$m) force amplitude and phase at the third
harmonic of the drive frequency ($3\omega$), for 119 $\mu$m of bimorph
amplitude and 31.5 $\mu$m edge-to-edge $z$-separation between masses with 6
$\mu$m tilt across the drive mass in the scanning ($y$) direction. The
horizontal axis shows the $y$-equilibrium position of the drive mass
oscillation. For reference, the $100$ $\mu$m-wide gold bars are centered in
the calculation at $y=-250$ $\mu$m, $y=-50$ $\mu$m, $y=150$ $\mu$m, and
$y=350$ $\mu$m, with silicon bars in between. \label{gravsim3w}}
\end{center}

\end{figure}

The amplitude and phase of either a magnetic or Yukawa signal will change
in a predictable way as we vary the $y$-equilibrium position of the drive
mass oscillation {\it{in situ}}.  The $x$,$y$, and $z$-separation between
the drive mass and test mass is controlled by a partially ($y$,$z$)
motorized manipulator stage at the top of the cryostat, and the tilt and
$x$,$y$, and $z$-separation between the masses are detected by capacitive
sensors \cite{stanford1,stanford2}. The expected Yukawa signal for a
typical vertical mass separation and bimorph amplitude is shown in Fig.
\ref{gravsim3w}. The vertical axis shows the calculated amplitude and phase
of a Yukawa ($\alpha=1000,\lambda=18$ $\mu$m) force at the third harmonic
of the drive (which equals the resonance frequency of the cantilever).  The
horizontal axis shows the $y$-equilibrium position of the oscillation. Note
that the third harmonic of the force (along with other odd harmonics)
vanishes when the equilibrium position of the drive mass is such that a
gold bar or silicon bar is centered underneath the cantilever. Conversely,
the force is maximized when the interface of the bars is centered beneath
the cantilever.  This results in a 100-$\mu$m spatial periodicity for the
amplitude of a Yukawa-type signal.

\section{Magnetic Calibration}\label{magcal}

Having a magnetic force for calibration that is switchable {\it{in situ}}
allows a great improvement in systematics by providing additional
information about the expected phase structure of a Yukawa-type signal and
by allowing multiple days of data to be analyzed together, as explained in
Sec. \ref{data}. When current is applied in the meandering pattern of gold
bars in the drive mass, the resulting magnetic field gradient above the
pattern couples to a magnetic moment on the cantilever to provide a
measurable force.  This allows the relative position of the test mass with
respect to the drive mass to be determined, as explained in the following
subsection. When the current is turned off, for the Yukawa-measurement to
have a clean background, it is necessary to eliminate or suppress the force
due to the differing magnetic susceptibility of the gold and silicon bars
in the drive mass when the drive mass becomes magnetized in an ambient
magnetic field such as the Earth's (see Sec. IIIB).

\subsection{Principle of the calibration}

\begin{figure}\begin{center}
\includegraphics[width=1.0 \columnwidth]{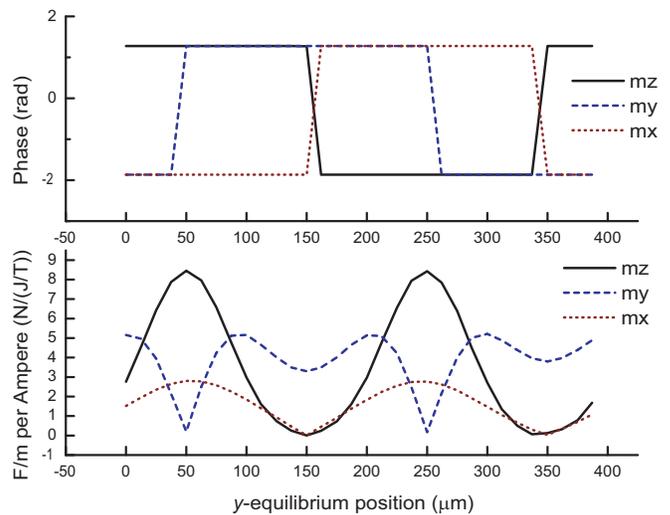}
\caption{(color online). Calculated magnetic force at the third harmonic of
the drive frequency, per unit magnetic moment per ampere of current through
the meander for varying $y$-equilibrium position of the drive mass
oscillation. Results are shown for 119 $\mu$m of bimorph amplitude and 30
$\mu$m face-to-face $z$-separation, for net magnetic moment in the $z$,$y$,
and $x$ directions, respectively. \label{magsim3w}}
\end{center}

\end{figure}

The vertical force on the cantilever with permanent magnetic moment
$\vec{m_c}$ in a magnetic field $\vec{B}$ is given by \begin{eqnarray} F_z
= (\vec{m_c} \cdot \vec{\nabla}) B_z. \label{magforce}
\end{eqnarray} Also a torque on the cantilever $\vec{\tau} = \vec{m_c}
\times \vec{B} $ can result in an effective force \be F^{\tau}_z =
\tau_z/l_{c} \label{magtorque} \ee where $l_c$ is the length of the
cantilever. Simulations are performed to model the magnetic field from
current flowing in the drive mass pattern, and the magnitude of the
magnetic moment can be deduced from Eqs. (\ref{magforce},\ref{magtorque})
and the measured force on the cantilever, provided the direction of the
magnetic moment is known.  If the direction of the net moment is uncertain,
only an approximate magnitude can be determined (see Fig. \ref{magsim3w}).

As we scan the $y$-equilibrium position of the oscillation {\it{in situ}},
 the magnetic signal at the third harmonic of the motion frequency
varies with a periodicity of 200 $\mu$m, corresponding to the spacing
between the gold bars.  In particular, as the $y$-equilibrium position of
the drive mass oscillation is scanned, the amplitude of the force on the
cantilever goes through a series of maxima and minima with a separation of
200 $\mu$m between adjacent maxima and minima.  Correspondingly, the phase
of the force on the cantilever experiences a shift of $\pi$ across each
magnetic minimum.  The particular locations of the maxima and minima with
respect to the position of the gold bars depends on the direction of the
net magnetic moment.  This is in contrast to the Yukawa signal, for which
the maxima occur when the equilibrium position of the drive mass
oscillation is such that the test mass is centered over the interface
between the gold and silicon bars.  The calculated magnetic force on the
cantilever per ampere of current in the drive mass per unit magnetic moment
is shown in Fig. \ref{magsim3w} for the cases of a net moment aligned in
the $x$,$y$, or $z$ directions, respectively.  The vertical separation
between the drive and test masses is taken to be 40 $\mu$m and the bimorph
amplitude is 119 $\mu$m.  The horizontal axis shows the $y$-equilibrium
position of the bimorph oscillation.

To obtain a proof-of-principle measurement of the magnetic response, an
early test was performed with a magnetic test mass consisting of a gold
rectangular prism coated with 100 nm of nickel \cite{stanford2}.  The
$y$-equilibrium position of the drive mass oscillation was varied across
the entire drive mass pattern as the force on the cantilever was studied
with and without current running through the drive mass meander. With
current in the drive mass meander, the force occurred with a 200 $\mu$m
periodicity as expected, corresponding to the spacing between the gold bars
carrying current.  With no current flowing across the drive mass, the force
occurred with a 100 $\mu$m periodicity and is due to the magnetization of
the drive mass in the ambient earth's magnetic field. It is necessary to
eliminate or suppress this susceptibility-dependent force in order to
detect a new non-Newtonian signal.

The phase of the magnetic force both with and without current flowing
across the drive mass is identical modulo $\pi$, as seen in earlier work
\cite{stanford2} and verified in simulation. Simulation indicates that the
phase of any Yukawa signal will also be identical to the phase of the
magnetic force modulo $\pi$. This is a useful handle in the identification
of a Yukawa force (see Sec. \ref{data}).

\subsection{Magnetic shield}

In order to minimize the effect of the Earth's field, we employ a
dual-layer magnetic shield assembly consisting of 0.050" thick high
magnetic permeability material provided by Amuneal Corporation.  The shield
is designed so it can be placed in position surrounding the cryostat. For
data collection the cryostat is hung from the ceiling and hangs freely
within the inner diameter of the magnetic shield. Numerical simulations
performed by Amuneal Corporation \cite{amuneal} indicated an expected
transverse shielding factor of 260 and longitudinal shielding factor of
160. At Stanford, the earth's magnetic field is primarily vertical with a
$z$-component of 45 $\mu$T, so the longitudinal figure is relevant. The
shield is degaussed prior to the experimental runs.

\subsection{Co/Pt multilayer films}

\begin{figure}\begin{center}
\includegraphics[width=1.0 \columnwidth]{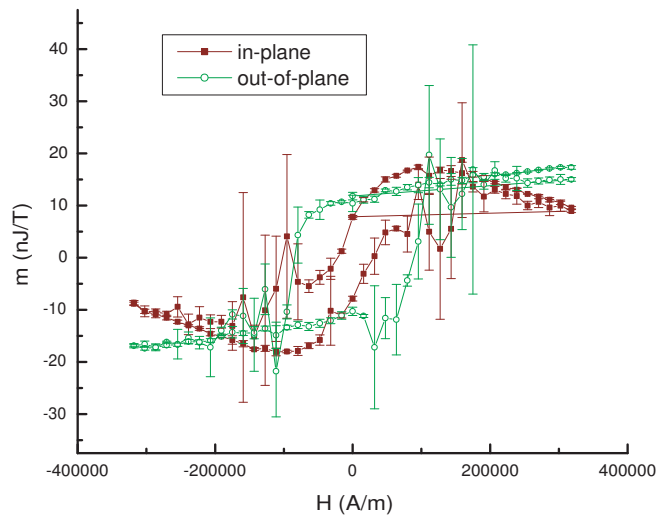}
\caption{(color online). Magnetic moment measured by MPMS-SQUID for
deposition of 3 ${\rm{\AA}}$ Co/ 15 ${\rm{\AA}}$ Pt/3 ${\rm{\AA}}$ Co/ 30
${\rm{\AA}}$ Pt on gold foil substrate, including seed layer of 100
${\rm{\AA}}$ Pt deposited at 350C, for in-plane and out-of-plane
magnetization. \label{foildep}}
\end{center}
\end{figure}

The ideal magnetic moment would have a known magnitude and direction.
Cobalt/platinum multi-layer films have been shown to have out-of-plane
anisotropy for limited ranges of cobalt and platinum thickness
\cite{bertero,clemens}.  In order to obtain out-of-plane anisotropy the
cobalt film thickness must be between 2.5 ${\rm{\AA}}$ and 11.5
${\rm{\AA}}$. The platinum spacing layers must also be larger than $\sim 8$
$ {\rm{\AA}}$ \cite{bertero}. For the test masses, 3 ${\rm{\AA}}$ Co/ 15
${\rm{\AA}}$ Pt/ 3 ${\rm{\AA}}$ Co/ 30 ${\rm{\AA}}$ Pt is sputtered on a 3
mm square gold foil substrate, on top of a seed layer of 100 ${\rm{\AA}}$
Pt deposited at 350 C. Increasing the platinum thickness from 11 to 15
${\rm{\AA}}$ resulted in improved out of plane anisotropy. The foil is then
inserted into an MPMS-SQUID \cite{squid}, and a hysteresis loop is studied
for the film magnetized both out-of-plane and in-plane for fields of -0.318
MA/m to 0.318 MA/m. The coercive field is around 0.056 MA/m and full
saturation occurs above 0.24 MA/m. The results shown in Fig. \ref{foildep}
indicate an out-of-plane easy axis, although the in-plane magnetization
loop also shows significant hysteresis.  We attribute this to the
imperfections on the foil surface as well as its curvature and roughness.
Due to the non-ideality, the data analysis of the experiment assumes that
the direction of the magnetic moment is an unknown parameter. However the
magnitude of the moment can be reasonably well controlled as discussed in
the following subsection.

\subsection{Magnetic Mass preparation}

The force due to the differing magnetic susceptibility of gold and silicon
depends on the ambient field $B_0$ which magnetizes the drive mass.  $B_0$
gets a contribution from the Earth's field, but also may be enhanced by the
local magnetic environment. In particular the magnetic field emanating from
the ferromagnetic layer on the test mass contributes to the drive mass
magnetization.   The differential volume-magnetic-susceptibility of gold
and silicon $\Delta \chi$ is $3 \times 10^{-5}$, leading to a differential
magnetization of order $\Delta M = \frac{1}{\mu_0} \Delta \chi B_0$.  The
correction to the local magnetic field and its gradient near the drive mass
surface can be estimated by integrating the induced magnetic dipole moment
per unit volume over the volume of the drive mass.  Very roughly, the
differential magnetic force experienced when the magnetic test mass is
above gold versus above silicon can be approximated as \be \Delta F \approx
m_c V_{\rm{eff}} \mu_0 \Delta M / r^4, \label{diff} \ee where $r$ is the
distance to the test mass magnetic film from the center of the drive mass
bar, $m_c$ is the magnetic moment on the cantilever, and $V_{\rm{eff}}$ is
the effective volume of the drive mass within range to contribute to the
signal.  To obtain a precise estimate of this force the net direction of
the magnetic moment on the cantilever must be known; otherwise only
order-of-magnitude estimates are possible. The magnetic measurements
described in Refs. \cite{stanford2,andythesis} can be explained by Eq.
(\ref{diff}) to within uncertainties, and therefore Eq. (\ref{diff}) can be
used to determine the target amount of magnetic material on the test mass.

To obtain a large enough force signal for a reasonable drive mass current
($10$ mA), while also having negligible magnetic susceptibility force with
the current off, it is sufficient to have a magnetic shield attenuation
factor greater than 100 and a total moment $m$ satisfying $5\times
10^{-15}$ J/T $< m <$ $5 \times 10^{-13}$ J/T.  For two layers of 3
${\rm{\AA}}$ Co / 15 ${\rm{\AA}}$ Pt, measurements suggest an area of (25
$\mu$m)$^2$ yields a moment of $5 \times 10^{-13}$ J/T.  The foil is
inserted into the FIB and sections of the magnetic layer near an edge of
the foil are removed with a 100 pA ion beam. 12 $\mu$m squares of magnetic
material are left in the locations that will become the centers of the test
masses and should result in a moment of approximately $1 \times 10^{-13}$
J/T.  A high-current 20 nA ion beam is then used to cut away the edge of
the foil to produce a clean edge.  The test masses are defined using the
high current beam. Care is taken to ensure that the high current beam does
not directly etch any magnetic material, as it has a tendency to deposit
material as the beam cuts.  A thin film of gold covers the magnetic patches
on the test mass due to this effect. The masses are completely free except
for a small ``tail''. At this point the masses are reinserted into the
SQUID and re-magnetized out-of-plane. This re-magnetization is performed to
eliminate the effects of any demagnetization that may have occurred from
heating while the high current beam cuts through the foil. The resolution
of the SQUID is insufficient to measure the magnetic moment of a single
test mass. However, the bulk foil can be seen to attain the same value of
magnetic moment as originally measured. After re-magnetization the masses
are mechanically extracted from the foil using non-magnetic tungsten probe
tips. In this way subsequent heating from the ion beam can be avoided. The
masses are epoxied onto the cantilever using Hysol-Loctite 1C epoxy
\cite{epoxypatch} and again only handled with a tungsten probe tip to avoid
magnetic contamination.

\section{Noise and Background Sources}\label{noise}

{\it{Thermal Noise.}} The minimum detectable force due to thermal noise at
temperature $T$ is \be F_{\rm{min}} = \sqrt{\frac{4 k_B T B k}{\omega_0
Q}}, \label{thn} \label{thnoise} \ee where $B$ is the bandwidth of the
measurement, $k$ is the cantilever spring constant, and $f_0 =
\omega_0/(2\pi)$ is its resonance frequency. This represents a fundamental
limit on the sensitivity of the measurement technique.

{\it{Vibrational Noise.}}  At the cantilever resonance frequency, assuming
a $4\%$ non-linearity in the bimorph and $Q$ of $10^5$, we estimate the
required internal vibration isolation (between the bimorph and cantilever)
to be $1.3 \times 10^5$, employing a simple analytic model
\cite{andythesis}.  Two sets of spring-mass stages isolate the platform on
which the bimorph is mounted from the mount holding the cantilever,
providing a factor of $10^6$ attenuation at $100$ Hz and $10^8$ attenuation
at $300$ Hz \cite{stanford2}.  The mass stages are also connected by thin
wires in addition to the springs to operate the various sensors and
piezo-electric stack. If these wires become too taut during the cooling of
the probe, the vibration isolation system can become compromised. A direct
measurement of the vibrational coupling at 300 Hz is unfortunately not
possible due to the operating frequency range of the bimorph.  The cryostat
is hung from $1$ Hz springs from a thick concrete ceiling to attenuate
unwanted external vibration at $300$ Hz.

{\it{Electrostatics.}}  The stainless steel frame which holds the
cantilever wafer is grounded to the main probe ground, so that together
with the metallized shield, a Faraday cage surrounds the cantilever.  Any
voltage difference $\Delta V$ built up between the cantilever and shield
will result in attractive force between them of magnitude $ F=\epsilon_0 A
(\Delta V)^2 / 2d^2 $ where $A$ is the area of the cantilever surface, and
$d$ is the separation between the cantilever and shield. Any built up
charge on the cantilever can produce an electrostatic force. In a
measurement using a similar shield, the shield motion at 300 Hz was found
to have an upper bound of approximately $\delta d \sim 1$ pm
\cite{jacthesis,sjsthesis}. Therefore voltage differences of 100 mV can
drive the cantilever above thermal noise. Alternatively, if a spurious
electrical voltage signal $V_0 \sin{\omega_0 t}$ with $V_0 > 100$ $\mu$V
leaks into the shield, the cantilever can be driven above thermal noise. In
either case the signal could not be mistaken for a new Yukawa force since
it would not vary with the spatial periodicity of the drive mass pattern.

{\it{Casimir force.}} There is a relatively large $(\sim 3 \times 10^{-16}$
N) Casimir force present at all times between the cantilever and shield
membrane.  If constant, it does not impede a force measurement at the
cantilever resonance frequency.  There is also a Casimir interaction
between the shield membrane and drive mass ground plane. The oscillating
bimorph can drive the shield membrane into motion, and due to piezo
non-linearity, can excite it at the cantilever resonance frequency.  The
shield membrane is sufficiently stiff so that the modulation of the Casimir
force on resonance does not excite the cantilever beyond the level of
thermal noise $\sim 10^{-18}$ N for our measurement bandwidth
\cite{stanford1,stanford2}. Even if the shield were driven enough to excite
the cantilever on resonance above thermal noise, the signal would not
exhibit the $y$-equilibrium position dependence that a Yukawa signal would
have, and therefore could not be erroneously interpreted as a new force.

One can also ask whether the shield membrane and drive mass ground plane
are sufficient to screen the {\it{direct}} Casimir force between the drive
mass and test mass, given the finite conductivity, plasma frequency, and
thickness of the gold coating.  In fact, the drive mass ground plane alone
is enough to prevent the differential Casimir force from the gold and
silicon drive mass sections from being transmitted at a detectable level.
Following Ref. \cite{lambrecht} we employ a reflection-based model for
computing the Casimir force between two metallic walls.  We obtain a
reduction factor $\eta_F$, (defined precisely in Ref. \cite{lambrecht})
which describes the reduction of the Casimir force at small separations and
for finite conductivity.  The parameter $\eta_F$ is defined as $F_C =
\eta_F F_P$ where $F_{P}$ is the perfect conductor result.  We take a gold
test mass of thickness 30 $\mu$m as one mirror, and consider the
differential Casimir force for the situations of the other mirror being 25
$\mu$m away and composed either of 0.1 $\mu$m thickness Au (corresponding
to the drive mass ground plane) or 100.1 $\mu$m thickness Au (corresponding
to the drive mass ground plane plus gold bar). We find that the difference
in $\eta_F$ is 0.0028, corresponding to a differential Casimir force of
$2.4 \times 10^{-20}$ N, allowing $\alpha=1$ to be probed at $\lambda=20$
$\mu$m, which is well below the sensitivity of the experiment. The shield
membrane attenuates this Casimir interaction even further, rendering the
effect negligible.

{\it{Magnetic background.}} Apart from those discussed in Sec.
\ref{magcal}, there are other mechanisms for a magnetic coupling that occur
with the drive mass spatial periodicity. These include the
susceptibility-induced interaction between the drive mass and the gold part
of the test mass and eddy currents produced in the drive mass as it
oscillates in the remnant ambient field. Also, any charge built up in the
silicon sections of the drive mass can generate a weak magnetic field as
the drive mass moves.  All of these mechanisms are too weak to produce a
measurable signal in this experiment.

{\it{Interferometer noise}.} Sources of electronic and optical noise in the
interferometer are discussed in Refs. \cite{jacthesis,sjsthesis}.
Electronic noise consists of shot noise from the laser, and to a lesser
extent Johnson noise from the 10 M$\Omega$ feedback resistor and
fluctuations in the laser current supply, and combines to produce typically
a few $\mu$V/Hz$^{1/2}$. Optical noise can be produced by stray reflections
in the interferometer as well as mechanical shaking of the fiber.
Modulation of the laser at frequencies over 100 MHz serves to reduce stray
reflections by reducing the coherence length of the laser
\cite{lasermod,rugarintf2}. Mechanical shaking of the fiber can mimic
cantilever displacement, along with any electrical coupling at harmonics of
the bimorph drive signal.  The amplitude of these noise sources is assessed
by measuring with the drive frequency off-resonance from a subharmonic.
However, this voltage noise is not expected to vary periodically as the
$y$-equilibrium position of the drive mass oscillation is scanned and thus
does not mimic a Yukawa-like force.

\section{Data}\label{data}

The general cooling-down procedure and data acquisition technique are
discussed at length in Refs. \cite{stanford2,sjsthesis}.  The resonance
frequency $f_0$ of the cantilever ranged from 324.082 to 324.136 Hz while
near base temperature during the course of the experimental run which is
the central subject of this paper. The base temperature of the probe ranged
from approximately $11-13$ K. The quality factor as measured using the
ring-down technique (see for example Ref. \cite{sjsthesis}) was
approximately 85,000. The spring constant $k$ of $0.0062$ N/m was inferred
from the measured resonance frequency and calculated mass of the test mass.
The effective temperature of the cantilever is deduced from $k$ and the
square of the amplitude spectrum near the cantilever thermal peak through
the equipartition theorem.  This effective noise temperature was typically
measured to be $20-25$ K at low laser power, about 10 K above the base
probe temperature, as was typically seen in previous work \cite{stanford2}.
The quality factor and effective temperature of the cantilever were both
found to depend on the incident laser power.  The laser power was reduced
below $1$ $\mu$W before temperature and quality factor stopped changing.
The low-temperature fringe height (as defined in Table \ref{syserr}) was
correspondingly reduced to approximately 250 mV. The voltage noise on the
interferometer, as assessed by tuning the piezo driving frequency 100 mHz
off resonance, was at the level of $10^{-18}$ N for the working fringe
height and quality factor. For the gravity force measurement, the quality
factor of the cantilever was adjusted with feedback \cite{stanford2} to
$9500-10000$, and the effective temperature with feedback was between $2-3$
K. The thermal noise limited force sensitivity was $\sim 200$
aN/$\sqrt{\rm{Hz}}$.

Prior to collecting data in search of non-Newtonian gravity (referred to
hereafter as `gravity data'), the approximate $x$- and $y$-position of the
drive mass with respect to the test mass is coarsely determined by using
the magnetic force signal, and it is verified that the test mass is
positioned over the central meandering region of the drive mass. Gravity
data are then collected as a function of the $y$-equilibrium position of
the drive mass oscillation for a given $z$-separation.  For each point of
gravity data collected, a magnetic force measurement is done at the same
location by applying current across the drive mass.  Then the magnetic
force is scanned in the $y$-direction to determine the $y$-position of the
closest magnetic minimum.  In this way the relative $y$-position between
gravity data points can be determined even in the presence of drifts in the
$y$-capacitance sensor reading of a given $y$-position that can occur over
the time scale of days, as demonstrated in earlier experimental runs with a
nickel magnetic layer on the test mass \cite{sjsthesis}. Having the
magnetic calibration (switchable {\it{in situ}}) thus permits multiple days
of data to be analyzed together, a significant improvement over previous
renditions of the experiment.

\begin{figure}\begin{center}
\includegraphics[width=1.0 \columnwidth]{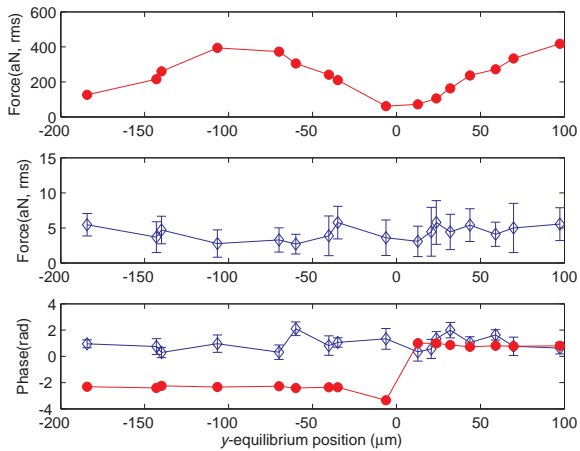}
\caption{(color online). Main data set showing statistical and systematic
uncertainties in the measured force. Diamonds show signal with current off.
Circles show signal with 10 mA current in the drive mass.  Lines are guides
to the eye. Error determination is described in Sec. \ref{analysis}. The
$y$-coordinates shown for the data set on the horizontal axis are the
displacements relative to one particular magnetic force minimum chosen as
$y=0$.  We note that these coordinates vary by an unknown offset with
respect to those shown in the calculations in Figs. \ref{gravsim3w} and
\ref{magsim3w}. \label{errorfig}}
\end{center}

\end{figure}

The data set is shown in Fig. \ref{errorfig}.  The magnitude and phase of
the magnetic (drive mass current on) and possible Yukawa (drive mass
current off) forces at the third harmonic are studied as the
$y$-equilibrium position of the drive mass oscillation is varied.  The 17
data records each consist of $24-120$ minutes of averaged data.  The amount
of averaging time chosen varies considerably due to the limited duty cycle
in the experiment. Environmental noise and possibly the internal
(bimorph-related) vibrational noise varied considerably over the course of
the data run.  By `environmental noise' we here refer to noise that is not
associated with the motion of the bimorph actuator.  In particular,
environmental noise may have come from the several nearby large scale
construction projects that were underway during various stages of the data
run. As a test of the environmental noise, measurements were taken with the
bimorph-actuator turned off.  Internal vibrational noise can in principle
be caused by a short of the vibration isolation springs due to taut probe
wiring. Although considerations are taken at room temperature when the
system is open to prevent taut wiring, it can be difficult to ensure that
no vibration isolation shorts develop after cooling.  The amount of
internal (bimorph-related) vibration noise can therefore vary based on the
$x$, $y$, and $z$-separation and tilts of the bimorph stage with respect to
the mass-stage holding the cantilever. The effects of vibration due to
internal isolation shorts were examined with the test and drive masses far
($\sim 1 $ mm) withdrawn from each other and the bimorph moving.  These
noise sources limited the amount of useful averaging time over the
experimental run.  The data in Fig. \ref{errorfig} were taken only at time
periods when such noise was minimal. Since we are not searching for
rare-event physics, it is legitimate to discard those other data for which
large disturbances were clearly present on the system.

The phase of the magnetic force undergoes a change of $\pi$ at the magnetic
force minimum at $y=0$.  If a pure Yukawa signal were present, one expects
to see two phase changes of $\pi$ in the drive mass current-off
(abbreviated hereafter as `current-off') data for every phase change in the
magnetic (current-on) signal, and two force minima for every magnetic force
minima. The $y$-extent of the data set should include three Yukawa phase
changes, each separated by 100 $\mu$m in the presence of such a force.  The
current-off data therefore shows no clear Yukawa signal in Fig.
\ref{errorfig}.  In fact, the phase of the current-off data remains
relatively constant over the data set.  In the presence of thermal noise
alone, the phase would be expected to be distributed randomly.  The bound
on any Yukawa signal associated with the data set is presented in Sec.
\ref{analysis}. The strength of the magnetic moment can be roughly
estimated from the current-on response. Simulation results indicate a
magnetic moment of approximately $1.3-3.7 \times 10^{-14}$ J/T, depending
on the direction of the moment.

The relatively constant phase of the current-off data could have a number
of possible origins. The driving signal applied to the piezo actuator could
have some small electronic leakage to the piezo near the cantilever. The
internal vibrational noise due to the bimorph oscillation can tend to occur
at a common phase given a common mechanism.  Bimorph non-linearity is also
responsible for some shield motion at the cantilever resonance frequency.
Although the associated modulation of the Casimir force is too small to be
observed, excess charge on the cantilever could couple to small
displacements of the shield membrane driven by the bimorph. The patch
effect can produce local potential variations in the gold coating on the
shield membrane which can vary with respect to the potential on the gold
test mass \cite{patcheffect}.  In any case, the expected 100 $\mu$m spatial
periodicity of a Yukawa-like signal can be distinguished from a background
force with a relatively constant phase. Therefore, an accurate bound on a
Yukawa-like interaction can be determined even in the presence of such a
background force, as discussed in Sec. \ref{analysis}.

\section{Error Analysis}\label{analysis}

The experimental error in the determination of either a bound on or
discovery of a Yukawa-type correction to Newtonian gravity can be sorted
into two types.  First there are the statistical and systematic
uncertainties in the measurements of the force $F_{\rm{meas}}$.  The second
type is the systematic error which enters the finite-element calculation of
the expected Yukawa force signal $F_Y$.   We treat the experimental
uncertainties in a Monte-Carlo fashion, based on the method described in
Ref. \cite{stanford2}. In order to obtain the function $\alpha (\lambda)$
for a Yukawa force that best fits the measured data we consider a
least-squares fit of the real and imaginary parts of the data to a set of
calculated forces in the Monte-Carlo simulation.   The four fitting
parameters, as a function of $\lambda$, are the result $\alpha$, an offset
variable $y_0$ to account for the lack of absolute $y$-position information
between the drive mass pattern and test mass due to the unknown direction
of the magnetic moment, and the real and imaginary parts of a constant
offset force ($R_0$, $I_0$). It is reasonable to include a constant offset
force to account for vibrational or other-type backgrounds that do not
depend periodically on the $y$-equilibrium position of the drive mass.

\subsection{Uncertainty in the measured force}

{\it{Statistical errors.}} The statistical error on each data point comes
primarily from thermal noise at the effective cantilever temperature and is
the most significant contribution $(90\% +)$ to the total 1$\sigma$ error
bars shown in Fig. \ref{errorfig}. The variation in the standard error
reflects the amount of averaging time for each data point, which varies
from 24 to 120 minutes. The statistical error decreases with the square
root of the numbers of samples.  There is a natural time scale for the
duration of statistically dependent samples given by the ring-down time of
the cantilever $\tau = \pi f_0 Q$.  We divide the data records into
subsections of length 30 seconds $\sim 3\tau$ and study the cross
correlation between these subsections to verify their statistical
independence, by the method discussed in Ref. \cite{stanford2}.

\begin{table}
\caption[]{Uncertainty in measured values of forces.  The voltage on the
interferometer is given by $ V_{\rm{intf}}(d) = V_c - \frac{V_{\rm{pp}}}{2}
\cos(4\pi d / \lambda)$, where $d$ is the distance between test mass and
the end of the optical fiber, and $\lambda$ is the laser wavelength. The
fringe height $V_{\rm{pp}} = (V_{\rm{max}}-V_{\rm{min}})$ and the fringe
center $V_{\rm{c}}$ is the mean of $V_{\rm{max}}$ and $V_{\rm{min}}$. The
fringe position is the deviation from the fringe center. \label{syserr}}
\vglue0.3cm
\begin{center}
\begin{tabular}{lcccc}
\hline \hline
& & & & \\
Parameter & Value & Error & Units & $dF\%$ \\
\hline
& & & & \\
Fringe height & .228-.472 & .002-.004  & V & 1  \\
Fringe position & 0 & 3 & \% & 0.2 \\
$f_{0}$ & 324.082-324.136 & .002 & Hz & 0.7  \\
$Q$ & 9500-10100 & 400 & - & 4.1 \\
$k$ & .0062 & .0003 & N/m & 4.5  \\
Fiber alignment & 0 & 22 & $\mu$m & 12  \\
 & & & \\
\hline
 & & & & \\
 Total & & & & 13 \\
 \hline\hline
\end{tabular}
\end{center}
\end{table}

{\it{Systematic errors in the measured force.}} The systematic errors in
the measured force are listed in Table \ref{syserr}. The error is dominated
by the uncertainty in the exact position of the fiber over the top surface
of the test mass. Considering the cantilever mode shape we estimate the
corresponding $1\sigma$ error in the measured force as $12\%$.  The
uncertainty including other systematics in the force measurement added in
quadrature is approximately $13\%$.

\subsection{Uncertainty in the calculated force}

{\it{Uncertainty in vertical separation and tilt.}} The calculated Yukawa
force strongly depends on the vertical separation between the drive and
test masses due to its exponential form.  The tilt of the drive mass with
respect to the shield wafer will also effectively increase the
$z$-separation, by an amount that depends on the $x$ and $y$ position of
the test mass with respect to the drive mass edges.
\begin{table}
\caption[]{Uncertainty in tilt and $z$-separation. Total tilts of the drive
mass relative to the shield plane are expressed in vertical distance over
the 1.8 mm length or 1.3 mm width of the drive mass.\label{tilts}}
\vglue0.3cm
\begin{center}
\begin{tabular}{lccc}
\hline \hline
& & &  \\
Parameter & Value & 1$\sigma$ Error & Units \\
\hline
& & &  \\
{\it{z-separation}} & & & \\
Shield-to-drive mass & 11-15 & 1  & $\mu$m  \\
Edge-edge between masses & 27-31 & 3 & $\mu$m  \\
Mean edge-edge separation & 29 & 3 & $\mu$m \\
MC Input  & 29 & 3 & $\mu$m \\
{\it{Relative to the shield plane}} & & & \\
 Total tilt in $y$-$z$ plane & 2 & 5 & $\mu$m/length  \\
 Total tilt in $x$-$z$ plane & 3 & 5 & $\mu$m/width \\
 & & &  \\
 \hline\hline
\end{tabular}
\end{center}
\end{table}
Table \ref{tilts} indicates the experimental uncertainty in the
$z$-separation and tilts. The tilts and $z$-separation are determined using
methods described in previous work \cite{stanford2}. From this table,
inputs to the Monte-Carlo are derived. The Monte-Carlo input for
$z$-separation is the mean edge-edge separation, adjusted to account for
the tilt in the drive mass silicon plane \cite{andythesis}. Variations in
$z$-separation between data points are very similar to variations within a
particular data point (1 $\mu$m) and typically are less than the overall
uncertainty in $z$-separation (3 $\mu$m). Also for the tilts, variation of
tilts across data points are much less than the overall uncertainty in the
tilt. Therefore for the Monte-Carlo we can treat the $z$-separation and
tilts of the entire data set as their mean value and add a Gaussian random
error with their standard deviation.  This is a reasonable approximation
since the standard deviation in the measured tilt or $z$-separation over
the 17 point data set is significantly less than the uncertainty in the
determination of the tilt or $z$-separation at any one data point. A more
detailed accounting of the determination of tilt and separation uncertainty
appears in Ref. \cite{andythesis}.

{\it{Uncertainty in bimorph amplitude.}} The mean bimorph amplitude
recorded for the data set was 119 $\mu$m with a standard deviation of 1.3
$\mu$m.  The $1\sigma$ uncertainty due to the motion calibration is 10
$\mu$m. The method for calibrating the bimorph and for determining the
error in bimorph amplitude is discussed at length in Ref. \cite{sjsthesis}.

{\it{Uncertainty in geometry and density of masses.}}
\begin{table}
\caption[]{Errors due to masses\label{multerr} } \vglue0.3cm
\begin{center}
\begin{tabular}{lc}
\hline \hline
& \\
Parameter & $d\alpha \%$ \\
\hline
&  \\
Volume of Test Mass & 4.5 \\
Voids in Drive Mass & 2.5 \\
Drive mass polished Au/Si boundary & 1 \\

 & \\
 \hline\hline
\end{tabular}
\end{center}
\end{table}
The uncertainty in the geometry of the drive mass is discussed at length in
Ref. \cite{sjsthesis}.  The uncertainty in the test mass geometry is
determined by a combination of optical and SEM imagery. Table \ref{multerr}
lists their relative contribution to the uncertainty in $\alpha$, which is
included as a multiplicative scaling error to the best fit results, using
the method discussed in Ref. \cite{stanford2}. The effective length, width,
and height shown in Table \ref{geoerr} are determined by considering the
approximately 84$^\circ$ sloping profile of the test mass walls measured by
the SEM and accounting for the extra mass present near the final-release
corner.  A SEM micrograph showing the mass attached to the cantilever after
being used in the data set is shown in the inset of Fig. \ref{schematic}.

\begin{table}
\caption[]{Uncertainty in Masses and Geometry\label{geoerr} } \vglue0.3cm
\begin{center}
\begin{tabular}{lccc}
\hline \hline
& & & \\
Parameter & Value & Error & Units  \\
\hline
& & &  \\
Test Mass length & 53 & 1 & $\mu$m   \\
Test Mass width & 54 & 1 & $\mu$m \\
Test Mass height & 27 & 1 & $\mu$m \\
Volume of Test Mass & 77000 & 3500 & $\mu$m$^3$ \\
Density of gold & 19300 & 0 & kg/m$^3$  \\
Total mass of test mass & 1.50 & 0.07 & $\mu$g  \\
 & & & \\
\hline
 & & &  \\
 \hline\hline
\end{tabular}
\end{center}
\end{table}

{\it{Uncertainty in relative $y$-position of data points.}} For the
Monte-Carlo analysis to be optimally constraining, the relative
$y$-separation of the data points must be known to within a few microns.
The $y$-coordinate of each data point has a $1 \sigma$ uncertainty of 2.5
$\mu$m. This accounts for the uncertainty in the $y$-position of the
magnetic minimum, as well as the uncertainty in the $y$-position at the
data collection point.

{\it{Uncertainty in the expected phase of a Yukawa-type signal.}}
Simulation indicates that the phase of any Yukawa signal will be identical
to the phase of the magnetic force modulo $\pi$.  The uncertainty in the
expected phase (modulo $\pi$) of a Yukawa signal is determined from the
uncertainty (0.1 rad) in the measured phase of the magnetic calibration
signal.

\subsection{Summary of inputs to the Monte-Carlo}\label{MCsec}

The set of geometrical inputs ($z$-separation and tilts in the $y$-$z$ and
$x$-$z$ planes, respectively) for the Monte-Carlo calculation are
summarized in Table \ref{tilts}. For each of the values of $\lambda$
considered (4,6,10,18,34,66 $\mu$m), the Yukawa force versus
$y$-displacement is calculated for 320 elements of a Gaussian distribution
of these geometrical parameters, with the means and standard deviations as
indicated.  For the mean value of bimorph amplitude, the corresponding
$3\omega$ force versus $y$-displacement is determined.  For each of these
curves, a least-squares fit to the measured forces and $y$-positions is
performed, yielding the best-fit $\alpha$, $y_0$, $R_0$, and $I_0$. The
measured force, measured $y$-position of the data points, and expected
phase of the Yukawa type signal are then `dithered' (with a Gaussian random
value added) 200 times according to the Gaussian distribution of their
statistical and systematic errors as described in the previous subsections.
For the uncertainty in phase, an element from a Gaussian distribution is
randomly assigned to the expected phase (modulo $\pi$) of a Yukawa type
signal, according to the uncertainty in the phase of the measured magnetic
signal. Each dithered set of data is then fit to yield a new best-fit value
of $\alpha$. This results in a set of 320 $\times$ 201 = 64,320
least-squares fits of the data to the calculated curves, producing 64,320
best-fit values of $\alpha$ for each $\lambda$. Finally, these values of
$\alpha$ are appropriately scaled to account for the uncertainty in the
bimorph amplitude and the uncertainty in the volume of the drive and test
masses, by using 50 samples taken from Gaussian distributions representing
these uncertainties. The treatment of these multiplicative errors is
further discussed in Ref. \cite{andythesis}. The end result is a histogram
of 3,216,000 best fit $\alpha$ for each of the six values of $\lambda$
explicitly considered. The numbers of samples are chosen to be sufficiently
large so that the means and 95-th percentiles of the histograms do not vary
significantly with the numbers of samples.

\section{Results}\label{results}

\begin{figure}\begin{center}
\includegraphics[width=1.0 \columnwidth]{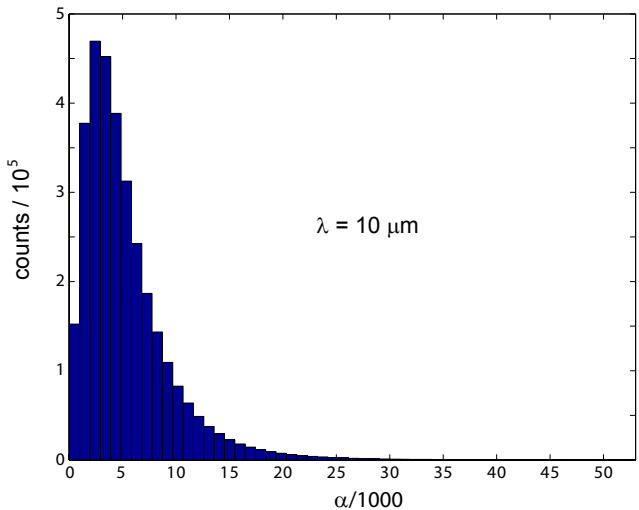}
\caption{(color online). Histogram of best-fit $\alpha$ results for
$\lambda = 10$ $\mu$m. \label{l10hist}}
\end{center}\end{figure}

\begin{figure}\begin{center}
\includegraphics[width=1.0 \columnwidth]{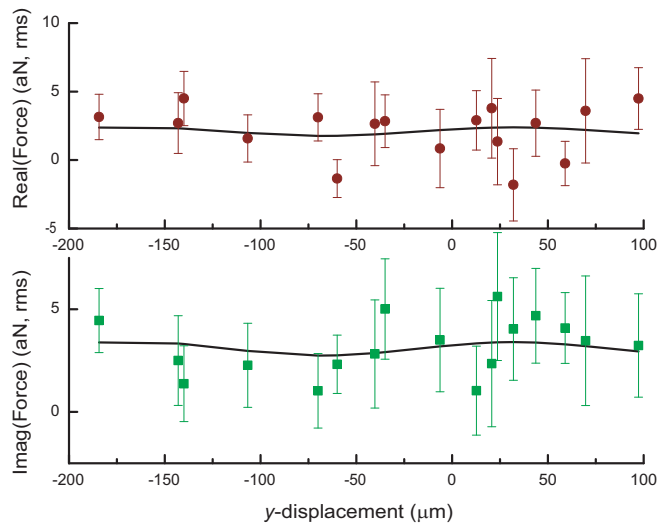}
\caption{(color online). Real and imaginary parts of measured force. A
typical Monte-Carlo fit result (best fit Yukawa signal plus offset) for
$\lambda = 10$ $\mu$m, $\alpha = 4450$, is shown as a solid line. Error
bars ($1\sigma$) on the measured forces are also shown. \label{realimag}}
\end{center}

\end{figure}

\begin{figure}\begin{center}
\includegraphics[width=1.0 \columnwidth]{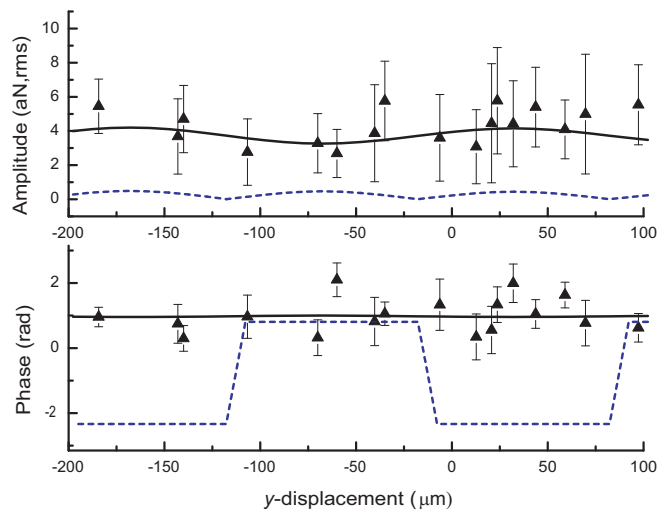}
\caption{(color online). Amplitude and phase of measured force. A typical
Monte-Carlo fit result (best fit Yukawa signal plus offset) for $\lambda =
10$ $\mu$m, $\alpha = 4450$, is shown as a solid line. Error bars
($1\sigma$) on the measured forces are also shown. Dotted line shows fit
result with best-fit offset-force subtracted.  The amplitude of the
best-fit offset-force is significantly larger than the amplitude of the
Yukawa component. \label{ampphase}}
\end{center}

\end{figure}

The histogram resulting from the Monte-Carlo analysis for $\lambda = 10$
$\mu$m is shown as an example in Fig. \ref{l10hist}. To illustrate a
typical fit of the force calculation to the data we consider an element
near the mean of the histogram for $\lambda$ of 10 $\mu$m, corresponding to
$\alpha = 4450$. Figure \ref{realimag} shows the real and imaginary parts
of the data along with statistical and systematic error bars. Also shown is
the best fit Monte-Carlo result for comparison. Figure \ref{ampphase} shows
the amplitude and phase of the data along with statistical and systematic
error bars.  The solid line shows the best fit Monte-Carlo result to the
data. We note that the actual fit is performed on the real and imaginary
parts of the data --- the amplitude and phase are plotted for illustration.
For reference, the dotted line indicates the best fit result after removing
the constant offset force. The fit is performed with the offset included
and the offset is subtracted afterwards. The data set is clearly best fit
with a constant offset force included, as evidenced by the concentration of
phase near 1 rad for the entire data set.  The fact that the best-fit phase
does not vary much reflects the adding of the best-fit Yukawa component
(with an amplitude of 0.5 aN) to the larger background best-fit offset rms
force vector (2.1 aN, 3.1 aN).

\begin{figure}\begin{center}
\includegraphics[width=1.0 \columnwidth]{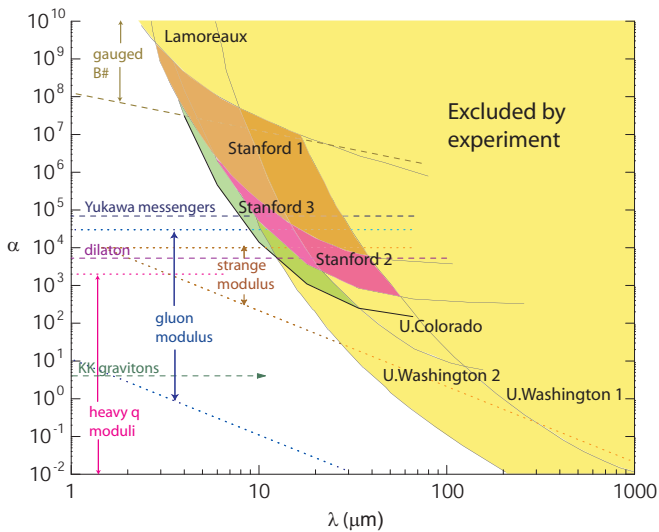}
\caption{(color online). New limits on Yukawa-forces.  The shaded area
refered to as `Stanford 3' represents the $95\%$ exclusion region yielded
with the present data set, listed in Table \ref{limits}. Experimental
limits derived in this work are on $|{\alpha}|$. The previous results
labeled `Stanford 1' and `Stanford 2' are presented in references
\cite{stanford1} and \cite{stanford2}, respectively. Also shown are
theoretical predictions and other experimental results as indicated in
Refs.
\cite{pricenature,adelberger,irvine,lamoreaux,andysavas,adelberger06_2}.
\label{resultfig2}}
\end{center}\end{figure}

The mean values of $\alpha$ and 95$\%$ confidence exclusion bounds for
Yukawa type forces are derived from the set of six histograms and listed in
Table \ref{limits}.  The ratio of the 95$\%$ confidence limit to the mean
value increases as $\lambda$ decreases, reflecting the increased
significance of the uncertainties in $z$-separation and tilt. The means of
each histogram represent the most likely value of any Yukawa force that may
be present.  Upcoming experiments will exhaustively re-examine this
parameter space. The results represent an improvement of the constraints on
Yukawa forces by approximately half an order-of-magnitude over our
previously published work \cite{stanford2}. We also note that the mean
values of $\alpha$ from the histograms in Ref. \cite{stanford2} are now
excluded at better than the 95$\%$ level by the present work.  Our final
results are shown in Fig. \ref{resultfig2}. We show only the $95\%$
confidence exclusion results from Table \ref{limits}.

\begin{table}
\caption[]{Experimental limits on Yukawa forces\label{limits}} \vglue0.3cm
\begin{center}
\begin{tabular}{ccc}
\hline \hline
& & \\
$\lambda$($\mu$m)   & Mean(MC) $\alpha$ $\space \space $  & 95\% exclusion $\alpha$ \\
\hline
& & \\
4 & $8.6 \times 10^6$ & $3.1 \times 10^7$  \\
6 & $1.6 \times 10^5$ & $4.6 \times 10^5$  \\
10 & $5.6 \times 10^3$ & $1.4 \times 10^4$  \\
18 & $5.1 \times 10^2$ & $1.1 \times 10^3$  \\
34 & $1.2 \times 10^2$ & $2.5 \times 10^2$  \\
66 & $7.0 \times 10^1$ & $1.5 \times 10^2$  \\
 & & \\
 \hline\hline
\end{tabular}
\end{center}
\end{table}

\section{Future Directions}\label{futurework2}

One limiting factor of the experimental run had to do with the duty cycle
of the experiment.  Improvements to the duty cycle could be made by adding
more automation to the experiment or transferring the apparatus to a
generally less noisy environment.  A substantial improvement could involve
the redesign of the cantilever to allow a larger area test mass. Such an
improvement is underway in a second-generation rotary-drive experiment
\cite{dmwthesis}, expected to be one to two orders of magnitude more
sensitive than the present apparatus. A number of other improvements could
be implemented resulting in more marginal gains. A doubly-clamped
cantilever beam loaded with a test mass in its center could reduce the
systematic error associated with the transverse position of the fiber above
the test mass. In this configuration, any transverse displacement of the
fiber along the test mass length affects the measured vertical displacement
to a lesser degree, as the cantilever mode shape is quadratic rather than
linear at the mass-attachment point. An additional piezo-electric
transducer that, unlike the bimorph, can be operated at the cantilever
resonance frequency could be installed near the bimorph to further
characterize the effectiveness of the vibration isolation. Voltage noise on
the interferometer could potentially be reduced by improving the vibration
isolation between the optical fiber in the cryostat and the bimorph
actuator.  A new set of shield wafers could be fabricated from a
double-polished silicon wafer to allow a sharper reflection of the drive
mass image during the room temperature alignment procedure.  This could
potentially reduce uncertainty in the tilt, resulting in decreased
effective separation of the masses.

\section{Conclusion}
To date, the gravitational interaction remains as one of the least
well-understood and least-tested aspects of fundamental physics.  In this
paper we have presented the latest results from the first-generation
Stanford micro-cantilever experiment, which in sum has improved the limits
on new Yukawa-type forces at 20 $\mu$m by over four orders of magnitude
since its conception.  Our most recent experimental constraints on
Yukawa-type deviations from Newtonian gravity represent the best bound in
the range of $5-15$ $\mu$m, with a $95\%$ confidence exclusion of forces
with $|\alpha|$ $>$ 14,000 at $\lambda=10$ $\mu$m.

\section{Acknowledgements}

We acknowledge many helpful discussions with Dan Rugar and Susan Holmes and
thank Savas Dimopoulos for providing much of the motivation for this work.
We thank the Bruce Clemens group for assistance with the Co/Pt depositions.
This work is supported by grant NSF-PHY-0554170.

    \bibliographystyle{apsrev} 
    \bibliography{frogland08prdv3}

\begin{thebibliography}{27}
\expandafter\ifx\csname natexlab\endcsname\relax\def\natexlab#1{#1}\fi
\expandafter\ifx\csname bibnamefont\endcsname\relax
  \def\bibnamefont#1{#1}\fi
\expandafter\ifx\csname bibfnamefont\endcsname\relax
  \def\bibfnamefont#1{#1}\fi
\expandafter\ifx\csname citenamefont\endcsname\relax
  \def\citenamefont#1{#1}\fi
\expandafter\ifx\csname url\endcsname\relax
  \def\url#1{\texttt{#1}}\fi
\expandafter\ifx\csname urlprefix\endcsname\relax\def\urlprefix{URL }\fi
\providecommand{\bibinfo}[2]{#2}
\providecommand{\eprint}[2][]{\url{#2}}

\bibitem[{\citenamefont{Dimopoulos and Guidice}(1996)}]{sg}
\bibinfo{author}{\bibfnamefont{S.}~\bibnamefont{Dimopoulos}} \bibnamefont{and}
  \bibinfo{author}{\bibfnamefont{G.~F.} \bibnamefont{Guidice}},
  \bibinfo{journal}{Phys.\ Lett.\ B} \textbf{\bibinfo{volume}{379}},
  \bibinfo{pages}{105} (\bibinfo{year}{1996}).

\bibitem[{\citenamefont{Antoniadis
  et~al.}(1998{\natexlab{a}})\citenamefont{Antoniadis, Dimopoulos, and
  Dvali}}]{add97}
\bibinfo{author}{\bibfnamefont{I.}~\bibnamefont{Antoniadis}},
  \bibinfo{author}{\bibfnamefont{S.}~\bibnamefont{Dimopoulos}},
  \bibnamefont{and} \bibinfo{author}{\bibfnamefont{G.}~\bibnamefont{Dvali}},
  \bibinfo{journal}{Nucl.\ Phys.\ B} \textbf{\bibinfo{volume}{516}},
  \bibinfo{pages}{70} (\bibinfo{year}{1998}{\natexlab{a}}).

\bibitem[{\citenamefont{Arkani-Hamed et~al.}(1998)\citenamefont{Arkani-Hamed,
  Dimopoulos, and Dvali}}]{add1}
\bibinfo{author}{\bibfnamefont{N.}~\bibnamefont{Arkani-Hamed}},
  \bibinfo{author}{\bibfnamefont{S.}~\bibnamefont{Dimopoulos}},
  \bibnamefont{and} \bibinfo{author}{\bibfnamefont{G.}~\bibnamefont{Dvali}},
  \bibinfo{journal}{Phys.\ Lett.\ B} \textbf{\bibinfo{volume}{429}},
  \bibinfo{pages}{263} (\bibinfo{year}{1998}).

\bibitem[{\citenamefont{Antoniadis
  et~al.}(1998{\natexlab{b}})\citenamefont{Antoniadis, Arkani-Hamed,
  Dimopoulos, and Dvali}}]{iadd}
\bibinfo{author}{\bibfnamefont{I.}~\bibnamefont{Antoniadis}},
  \bibinfo{author}{\bibfnamefont{N.}~\bibnamefont{Arkani-Hamed}},
  \bibinfo{author}{\bibfnamefont{S.}~\bibnamefont{Dimopoulos}},
  \bibnamefont{and} \bibinfo{author}{\bibfnamefont{G.}~\bibnamefont{Dvali}},
  \bibinfo{journal}{Phys.\ Lett.\ B} \textbf{\bibinfo{volume}{436}},
  \bibinfo{pages}{257} (\bibinfo{year}{1998}{\natexlab{b}}).

\bibitem[{\citenamefont{Arkani-Hamed et~al.}(1999)\citenamefont{Arkani-Hamed,
  Dimopoulos, and Dvali}}]{add3}
\bibinfo{author}{\bibfnamefont{N.}~\bibnamefont{Arkani-Hamed}},
  \bibinfo{author}{\bibfnamefont{S.}~\bibnamefont{Dimopoulos}},
  \bibnamefont{and} \bibinfo{author}{\bibfnamefont{G.}~\bibnamefont{Dvali}},
  \bibinfo{journal}{Phys.\ Rev.\ D} \textbf{\bibinfo{volume}{59}},
  \bibinfo{pages}{086004} (\bibinfo{year}{1999}).

\bibitem[{\citenamefont{Chiaverini et~al.}(2003)\citenamefont{Chiaverini,
  Smullin, Geraci, Weld, and Kapitulnik}}]{stanford1}
\bibinfo{author}{\bibfnamefont{J.}~\bibnamefont{Chiaverini}},
  \bibinfo{author}{\bibfnamefont{S.~J.} \bibnamefont{Smullin}},
  \bibinfo{author}{\bibfnamefont{A.~A.} \bibnamefont{Geraci}},
  \bibinfo{author}{\bibfnamefont{D.~M.} \bibnamefont{Weld}}, \bibnamefont{and}
  \bibinfo{author}{\bibfnamefont{A.}~\bibnamefont{Kapitulnik}},
  \bibinfo{journal}{Phys.\ Rev.\ Lett.} \textbf{\bibinfo{volume}{90}},
  \bibinfo{pages}{151101} (\bibinfo{year}{2003}).

\bibitem[{\citenamefont{Smullin et~al.}(2005)\citenamefont{Smullin, Geraci,
  Weld, Chiaverini, Holmes, and Kapitulnik}}]{stanford2}
\bibinfo{author}{\bibfnamefont{S.~J.} \bibnamefont{Smullin}},
  \bibinfo{author}{\bibfnamefont{A.~A.} \bibnamefont{Geraci}},
  \bibinfo{author}{\bibfnamefont{D.~M.} \bibnamefont{Weld}},
  \bibinfo{author}{\bibfnamefont{J.}~\bibnamefont{Chiaverini}},
  \bibinfo{author}{\bibfnamefont{S.}~\bibnamefont{Holmes}}, \bibnamefont{and}
  \bibinfo{author}{\bibfnamefont{A.}~\bibnamefont{Kapitulnik}},
  \bibinfo{journal}{Phys.\ Rev.\ D} \textbf{\bibinfo{volume}{72}},
  \bibinfo{pages}{122001} (\bibinfo{year}{2005}).

\bibitem[{\citenamefont{Smullin}(2005)}]{sjsthesis}
\bibinfo{author}{\bibfnamefont{S.}~\bibnamefont{Smullin}}, Ph.D. thesis,
  \bibinfo{school}{Stanford University} (\bibinfo{year}{2005}).

\bibitem[{\citenamefont{Rugar et~al.}(1989)\citenamefont{Rugar, Mamin, and
  Guethner}}]{rugarintf}
\bibinfo{author}{\bibfnamefont{D.}~\bibnamefont{Rugar}},
  \bibinfo{author}{\bibfnamefont{H.~J.} \bibnamefont{Mamin}}, \bibnamefont{and}
  \bibinfo{author}{\bibfnamefont{P.}~\bibnamefont{Guethner}},
  \bibinfo{journal}{Appl.\ Phys.\ Lett.} \textbf{\bibinfo{volume}{55}},
  \bibinfo{pages}{2588} (\bibinfo{year}{1989}).

\bibitem[{\citenamefont{Geraci}(2007)}]{andythesis}
\bibinfo{author}{\bibfnamefont{A.}~\bibnamefont{Geraci}}, Ph.D. thesis,
  \bibinfo{school}{Stanford University} (\bibinfo{year}{2007}).

\bibitem[{\citenamefont{Chiaverini}(2002)}]{jacthesis}
\bibinfo{author}{\bibfnamefont{J.}~\bibnamefont{Chiaverini}}, Ph.D. thesis,
  \bibinfo{school}{Stanford University} (\bibinfo{year}{2002}).

\bibitem[{\citenamefont{{Amuneal Manufacturing Corp.}}()}]{amuneal}
\bibinfo{author}{\bibnamefont{{Amuneal Manufacturing Corp.}}},
  \bibinfo{note}{4737 Darrah Street Philadelphia, PA 19124, USA}.

\bibitem[{\citenamefont{Bertero}(1996)}]{bertero}
\bibinfo{author}{\bibfnamefont{G.}~\bibnamefont{Bertero}}, Ph.D. thesis,
  \bibinfo{school}{Stanford University} (\bibinfo{year}{1996}).

\bibitem[{\citenamefont{{Bruce Clemens}}()}]{clemens}
\bibinfo{author}{\bibnamefont{{Bruce Clemens}}}, \bibinfo{note}{private
  communication}.

\bibitem[{\citenamefont{{Quantum Design, Inc.}}()}]{squid}
\bibinfo{author}{\bibnamefont{{Quantum Design, Inc.}}},
  \bibinfo{note}{\url{www.qdusa.com}, {Magnetic Property Measurement System}}.

\bibitem[{\citenamefont{{Epoxi-Patch, Dexter Corporation}}()}]{epoxypatch}
\bibinfo{author}{\bibnamefont{{Epoxi-Patch, Dexter Corporation}}},
  \bibinfo{note}{1 Dexter Dr., Seabrook, NH 03874, USA}.

\bibitem[{\citenamefont{Lambrecht and Reynaud}(2000)}]{lambrecht}
\bibinfo{author}{\bibfnamefont{A.}~\bibnamefont{Lambrecht}} \bibnamefont{and}
  \bibinfo{author}{\bibfnamefont{S.}~\bibnamefont{Reynaud}},
  \bibinfo{journal}{Eur.\ Phys.\ J.\ D} \textbf{\bibinfo{volume}{8}},
  \bibinfo{pages}{309} (\bibinfo{year}{2000}).

\bibitem[{\citenamefont{Kassies et~al.}(2004)\citenamefont{Kassies, van~der
  Werf, Bennink, and Otto}}]{lasermod}
\bibinfo{author}{\bibfnamefont{R.}~\bibnamefont{Kassies}},
  \bibinfo{author}{\bibfnamefont{K.~O.} \bibnamefont{van~der Werf}},
  \bibinfo{author}{\bibfnamefont{M.~L.} \bibnamefont{Bennink}},
  \bibnamefont{and} \bibinfo{author}{\bibfnamefont{C.}~\bibnamefont{Otto}},
  \bibinfo{journal}{Rev.\ Sci.\ Instrum.} \textbf{\bibinfo{volume}{75}},
  \bibinfo{pages}{689} (\bibinfo{year}{2004}).

\bibitem[{\citenamefont{Albrecht et~al.}(1992)\citenamefont{Albrecht,
  Gr\"{u}tter, Rugar, and Smith}}]{rugarintf2}
\bibinfo{author}{\bibfnamefont{T.~R.} \bibnamefont{Albrecht}},
  \bibinfo{author}{\bibfnamefont{P.}~\bibnamefont{Gr\"{u}tter}},
  \bibinfo{author}{\bibfnamefont{D.}~\bibnamefont{Rugar}}, \bibnamefont{and}
  \bibinfo{author}{\bibfnamefont{D.~P.~E.} \bibnamefont{Smith}},
  \bibinfo{journal}{Ultramicroscopy} \textbf{\bibinfo{volume}{42-44}},
  \bibinfo{pages}{1638} (\bibinfo{year}{1992}).

\bibitem[{\citenamefont{Speake and Trenkel}(2003)}]{patcheffect}
\bibinfo{author}{\bibfnamefont{C.~C.} \bibnamefont{Speake}} \bibnamefont{and}
  \bibinfo{author}{\bibfnamefont{C.~T.} \bibnamefont{Trenkel}},
  \bibinfo{journal}{Phys.\ Rev.\ Lett.} \textbf{\bibinfo{volume}{90}},
  \bibinfo{pages}{160403} (\bibinfo{year}{2003}).

\bibitem[{\citenamefont{Long et~al.}(2003)\citenamefont{Long, Chan, Churnside,
  Gulbis, Varney, and Price}}]{pricenature}
\bibinfo{author}{\bibfnamefont{J.~C.} \bibnamefont{Long}},
  \bibinfo{author}{\bibfnamefont{H.~W.} \bibnamefont{Chan}},
  \bibinfo{author}{\bibfnamefont{A.~B.} \bibnamefont{Churnside}},
  \bibinfo{author}{\bibfnamefont{E.~A.} \bibnamefont{Gulbis}},
  \bibinfo{author}{\bibfnamefont{M.~C.~M.} \bibnamefont{Varney}},
  \bibnamefont{and} \bibinfo{author}{\bibfnamefont{J.~C.} \bibnamefont{Price}},
  \bibinfo{journal}{Nature} \textbf{\bibinfo{volume}{421}},
  \bibinfo{pages}{922} (\bibinfo{year}{2003}).

\bibitem[{\citenamefont{Hoyle et~al.}(2004)\citenamefont{Hoyle, Kapner, Heckel,
  Adelberger, Gundlach, Schmidt, and Swanson}}]{adelberger}
\bibinfo{author}{\bibfnamefont{C.~D.} \bibnamefont{Hoyle}},
  \bibinfo{author}{\bibfnamefont{D.~J.} \bibnamefont{Kapner}},
  \bibinfo{author}{\bibfnamefont{B.~R.} \bibnamefont{Heckel}},
  \bibinfo{author}{\bibfnamefont{E.~G.} \bibnamefont{Adelberger}},
  \bibinfo{author}{\bibfnamefont{J.~H.} \bibnamefont{Gundlach}},
  \bibinfo{author}{\bibfnamefont{U.}~\bibnamefont{Schmidt}}, \bibnamefont{and}
  \bibinfo{author}{\bibfnamefont{H.~E.} \bibnamefont{Swanson}},
  \bibinfo{journal}{Phys.\ Rev.\ D} \textbf{\bibinfo{volume}{70}},
  \bibinfo{pages}{042004} (\bibinfo{year}{2004}).

\bibitem[{\citenamefont{Hoskins et~al.}(1985)\citenamefont{Hoskins, Newman,
  Spero, and Schultz}}]{irvine}
\bibinfo{author}{\bibfnamefont{J.~K.} \bibnamefont{Hoskins}},
  \bibinfo{author}{\bibfnamefont{R.~D.} \bibnamefont{Newman}},
  \bibinfo{author}{\bibfnamefont{R.}~\bibnamefont{Spero}}, \bibnamefont{and}
  \bibinfo{author}{\bibfnamefont{J.}~\bibnamefont{Schultz}},
  \bibinfo{journal}{Phys. Rev. D} \textbf{\bibinfo{volume}{32}},
  \bibinfo{pages}{3084} (\bibinfo{year}{1985}).

\bibitem[{\citenamefont{Lamoreaux}(1997)}]{lamoreaux}
\bibinfo{author}{\bibfnamefont{S.~K.} \bibnamefont{Lamoreaux}},
  \bibinfo{journal}{Phys.\ Rev.\ Lett.} \textbf{\bibinfo{volume}{78}},
  \bibinfo{pages}{5} (\bibinfo{year}{1997}).

\bibitem[{\citenamefont{Dimopoulos and Geraci}(2003)}]{andysavas}
\bibinfo{author}{\bibfnamefont{S.}~\bibnamefont{Dimopoulos}} \bibnamefont{and}
  \bibinfo{author}{\bibfnamefont{A.~A.} \bibnamefont{Geraci}},
  \bibinfo{journal}{Phys.\ Rev.\ D} \textbf{\bibinfo{volume}{68}},
  \bibinfo{pages}{124021} (\bibinfo{year}{2003}).

\bibitem[{\citenamefont{Kapner et~al.}(2007)\citenamefont{Kapner, Cook,
  Adelberger, Gundlach, Heckel, Hoyle, and Swanson}}]{adelberger06_2}
\bibinfo{author}{\bibfnamefont{D.~J.} \bibnamefont{Kapner}},
  \bibinfo{author}{\bibfnamefont{T.~S.} \bibnamefont{Cook}},
  \bibinfo{author}{\bibfnamefont{E.~G.} \bibnamefont{Adelberger}},
  \bibinfo{author}{\bibfnamefont{J.~H.} \bibnamefont{Gundlach}},
  \bibinfo{author}{\bibfnamefont{B.~R.} \bibnamefont{Heckel}},
  \bibinfo{author}{\bibfnamefont{C.~D.} \bibnamefont{Hoyle}}, \bibnamefont{and}
  \bibinfo{author}{\bibfnamefont{H.~E.} \bibnamefont{Swanson}},
  \bibinfo{journal}{Phys.\ Rev.\ Lett.} \textbf{\bibinfo{volume}{98}},
  \bibinfo{eid}{021101} (\bibinfo{year}{2007}).

\bibitem[{\citenamefont{Weld et~al.}()\citenamefont{Weld, Xia, Cabrera, and
  Kapitulnik}}]{dmwthesis}
\bibinfo{author}{\bibfnamefont{D.~M.} \bibnamefont{Weld}},
  \bibinfo{author}{\bibfnamefont{J.}~\bibnamefont{Xia}},
  \bibinfo{author}{\bibfnamefont{B.}~\bibnamefont{Cabrera}}, \bibnamefont{and}
  \bibinfo{author}{\bibfnamefont{A.}~\bibnamefont{Kapitulnik}},
  \bibinfo{note}{in press}.

\end{thebibliography}

\end{document}